\documentclass[aps,prb,twocolumn]{revtex4}
\usepackage{amssymb}
\usepackage{graphicx}

\begin{document}
\title{Antiferromagnetic Mott insulating state in single crystals of the hexagonal lattice material Na$_2$IrO$_3$}
\author{Yogesh Singh and P. Gegenwart}
\affiliation{I. Physikalisches Institut, Georg-August-Universit\"at G\"ottingen, D-37077, G\"ottingen, Germany}
\date{\today}

\begin{abstract}
We have synthesized single crystals of Na$_2$IrO$_3$ and studied their structure, transport, magnetic, and thermal properties using powder x-ray diffraction (PXRD), electrical resistivity, isothermal magnetization $M$ versus magnetic field $H$, magnetic susceptibility $\chi$ versus temperature $T$, and heat capacity $C$ versus $T$ measurements.  Na$_2$IrO$_3$ crystallizes in the monoclinic \emph{C2/c} (No.~15) type structure which is made up of Na and NaIr$_2$O$_6$ layers alternately stacked along the $c$ axis.  The $\chi(T)$ data show Curie-Weiss behavior at high $T > 200$~K with an effective moment $\mu_{\rm eff} = 1.82(1) \mu_{\rm B}$ indicating an effective spin $S_{\rm eff} = 1/2$ on the Ir$^{4+}$ moments.  A large Weiss temperature $\theta = - 116(3)$~K indicates substantial antiferromagnetic interactions between these $S_{\rm eff} = 1/2$, Ir$^{4+}$ moments.  Sharp anomalies in $\chi(T)$ and $C(T)$ data indicate that Na$_2$IrO$_3$ undergoes a transition into a long-range antiferromagnetically ordered state below $T_{\rm N} = 15$~K\@.  The magnetic entropy at $T_{\rm N}$ is only about 20\% of what is expected for $S_{\rm eff} = 1/2$ moment ordering.  The reduced entropy and the small ratio $T_N/\theta \approx 0.13$ suggest geometrical magnetic frustration and/or low-dimensional magnetic interactions in Na$_2$IrO$_3$.  In plane resistivity measurements show insulating behavior.  This together with the local moment magnetism indicates that bulk Na$_2$IrO$_3$ is a Mott insulator.

\end{abstract}
\pacs{75.40.Cx, 75.50.Lk, 75.10.Jm, 75.40.Gb}

\maketitle

\section{Introduction}
\label{sec:INTRO}
Layered transition metal oxides provide a playground for various strongly correlated phenomena.  The antiferromagnetic Mott insulating ground state in the spin $S = 1/2$ square lattice material La$_2$CuO$_4$ and the related high temperature superconductivity in doped materials,\cite{Johnston1997} spin-triplet superconductivity in Sr$_2$RuO$_4$,\cite{Nelson2004} metamagnetism and quantum criticality in Sr$_3$Ru$_2$O$_7$,\cite{Borzi2007} the metal-insulator transition in Cd$_2$Os$_2$O$_7$,\cite{Mandrus2001} and superconductivity in water intercalated Na$_x$CoO$_2$,\cite{Takada2003} are just some examples of these correlated behaviors.       
\begin{figure}[t]
\includegraphics[width=1.85in, angle=-90]{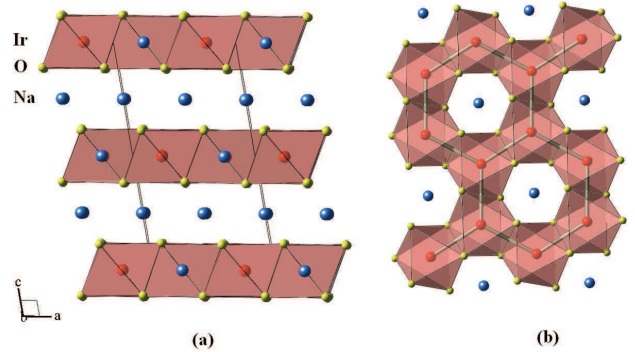}
\caption{(Color online) The crystallographic structure of Na$_2$IrO$_3$.  The Na, Ir, and O atoms are shown as blue, red, and yellow spheres, respectively.  (a) The view perpendicular to the $c$ axis showing the layered structure with layers containing only Na atoms alternating slabs of NaIr$_2$O$_6$ stacked along the $c$ axis.  The IrO$_6$ octahedra are shown in pink with the (red) Ir atoms sitting in the middle.  (b)  One of the NaIr$_2$O$_6$ slabs viewed down the $c$ axis to highlight the honeycomb lattice of Ir atoms within the layer.  The Na atoms occupy voids between the IrO$_6$ octahedra. 
\label{Fig-structure}}
\end{figure}

Electronic correlations are expected to be strongest in 3$d$ transition metals and are expected to decrease as one goes to 4$d$, and 5$d$ transition metals as the extent of the $d$ orbital increases.  Thus electronic correlations are expected to be weakest in 5$d$ materials and these materials are expected to be metallic due to the larger spatial extent of their $d$ orbitals.  However, recently several 5$d$ transition metal oxides like Sr$_2$IrO$_4$,\cite{Crawford1994} Sr$_3$Ir$_2$O$_7$,\cite{Cao2002} Ba$_2$NaOsO$_6$,\cite{Erickson2007} have rather surprisingly been discovered to show insulating behaviors.  As one goes from 3$d$ to 5$d$ transition metals the spin-orbit interaction also increases which leads to an effective angular momentum $J_{\rm eff}$ being a good quantum number as opposed to just the spin $S$.  The insulating state in Sr$_2$IrO$_4$ has in fact been suggested to be a novel Mott-insulating ground state arising from electron correlations among $J_{\rm eff} = 1/2$  Ir$^{4+}$ moments.\cite{Kim2008}  Thus in 5$d$ systems, the spin-orbit interactions and electron correlations are of comparable strength. 
Recently a new layered iridate Na$_2$IrO$_3$ has been studied theoretically and has been suggested to be a $J_{\rm eff} = 1/2$ system arising from strong spin-orbit interactions.\cite{Shitade2009}  Correlations between these effective spins has been predicted to lead to an antiferromagnetic insulating state.\cite{Shitade2009, Jin2009}  The hexagonal arrangement of the Ir$^{4+}$ moments in Na$_2$IrO$_3$ was also suggested to be a realization of the Kane-Mele model which was put forward for quantum spin Hall (QSH) effect in the honeycomb lattice of Graphene.\cite{Kane2005a,Kane2005b}  Thus Na$_2$IrO$_3$ was proposed to be a topological insulator and a possible candidate to show the QSH effect at room temperature.\cite{Shitade2009}
Recently $A_2$IrO$_3$ ($A =$~Li, Na) materials have also been suggested\cite{Chaloupka2010} to be experimental realization of the Kitaev model of spins $S = 1/2$ sitting on a hexagonal lattice.\cite{Kitaev2006}  The model consists of highly frustrated in-plane magnetic interactions which can lead to the spin liquid ground state.
Despite enormous theoretical interest however, to the best of our knowledge, no experimental information is reported for this material. 

Herein we report synthesis, structure, electrical transport, magnetic, and thermal properties of single crystalline Na$_2$IrO$_3$.     

\section{Experimental Details}
\label{sec:EXPT}
Polycrystalline samples of Na$_2$IrO$_3$ and Na$_2$SnO$_3$ were synthesized by solid state synthesis.  The starting materials Na$_2$CO$_3$ (99.995\% Alfa Aesar) and anhydrous IrO$_2$ ($\geq 99.95\%$ KaiDa) or SnO$_2$ (99.995\% Alfa Aesar) were mixed in the ratio 1.05 : 1 and placed in an alumina crucible with a lid and heated to 750~$^{\circ}$C in 5~hrs and held there for 24~hrs after which it was furnace cooled to room temperature.  The resulting mixture was ground and pelletized, placed in an alumina crucible and heated to 900~$^{\circ}$C in 6~hrs and held there for 48~hrs before cooling to room temperature.  The polycrystalline sample of Na$_2$SnO$_3$ was re-ground, pelletized and given a further heat treatment at 1000~$^{\circ}$C for 48~hrs.  

To grow single crystals of Na$_2$IrO$_3$, the above pellet of Na$_2$IrO$_3$ obtained at 900~$^{\circ}$C was reground and pelletized and heated to 1050~~$^{\circ}$C in 5~hrs, kept at this temperature for 72~hrs, slowly (10~$^{\circ}$C/hr) cooled to 900~$^{\circ}$C and then quenched in air.  Shiny plate-like single crystals of typical dimensions $1.5 \times 1.2 \times 0.1$~mm$^3$ were found to grow on top of a semi-melted pellet.  To optimize the growth, attempts to grow crystals similarly at temperatures between 1000--1150~~$^{\circ}$C were attempted.  Crystals were not found to grow below about 1000~~$^{\circ}$C and above about 1100~~$^{\circ}$C\@.  Thus, $T = 1050$~~$^{\circ}$C seems to be the optimal growth temperature.  The structure and composition of the polycrystalline and single crystalline Na$_2$IrO$_3$ samples were analyzed using powder x-ray diffraction (XRD) and chemical analysis using energy dispersive x-ray (EDX) analysis using a Philips scanning electron microscope (SEM).  The XRD patterns were obtained at room temperature using a Rigaku Geigerflex diffractometer with Cu K$\alpha$ radiation, in the 2$\theta$ range from 10 to 90$^\circ$ with a 0.02$^\circ$ step size. Intensity data were accumulated for 5~s per step.  Dc electrical transport was measured on a home built setup using the four probe technique.  The isothermal magnetization $M(H)$, and static magnetic susceptibility $\chi(T)$ were measured using a commercial Superconducting Quantum Interference Device (SQUID) magnetometer (MPMS5, Quantum Design) and the heat capacity $C(T)$ was measured using a commercial Physical Property Measurement System (PPMS5, Quantum Design).  

The magnetic measurements were performed on two kinds of samples.  The $M(H)$ and $\chi(T)$ were measured on a collection of randomly oriented crystals with total mass $m = 86.34$~mg and the anisotropic $\chi(T)$ data were measured on a collection of $6$ co-aligned crystals of total mass $m = 8.54$~mg.  The heat capacity $C(T)$ was measured on a collection of 4 crystals of total mass $m = 6.83$~mg.   
The magnetic and thermal properties of the polycrystalline Na$_2$IrO$_3$ synthesized at 900~$^{\circ}$C were dominated by disorder effects showing a spin-glass like behavior at low temperatures.  Therefore, physical properties of only the high quality single crystals are reported in the following as they most likely represent the intrinsic behavior of Na$_2$IrO$_3$.  The polycrystalline sample, along with the single crystals, is used only for structural characterization.

\begin{figure}[t]
\includegraphics[width=2.5in,angle=-90]{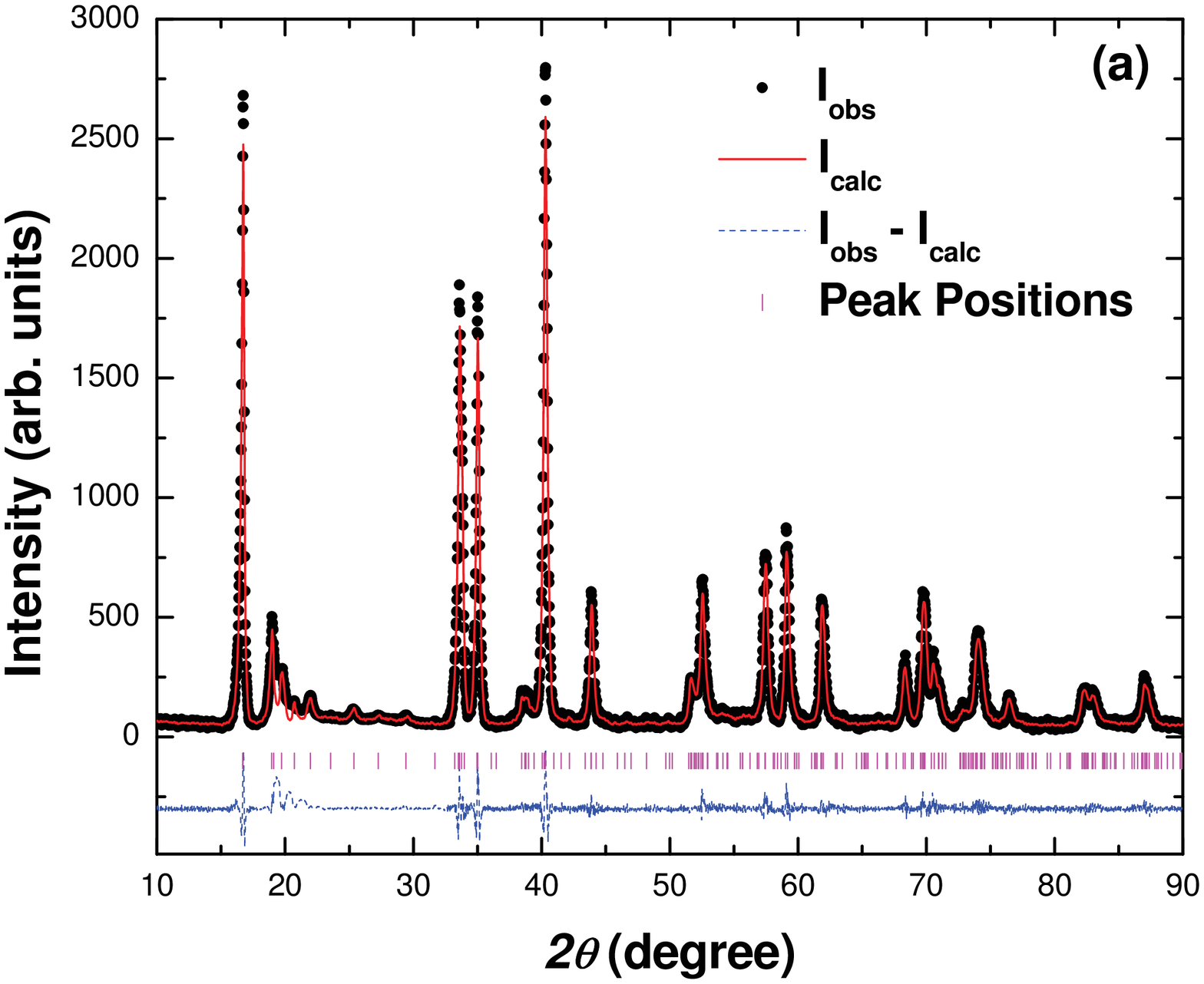}
\includegraphics[width=2.35in,angle=-90]{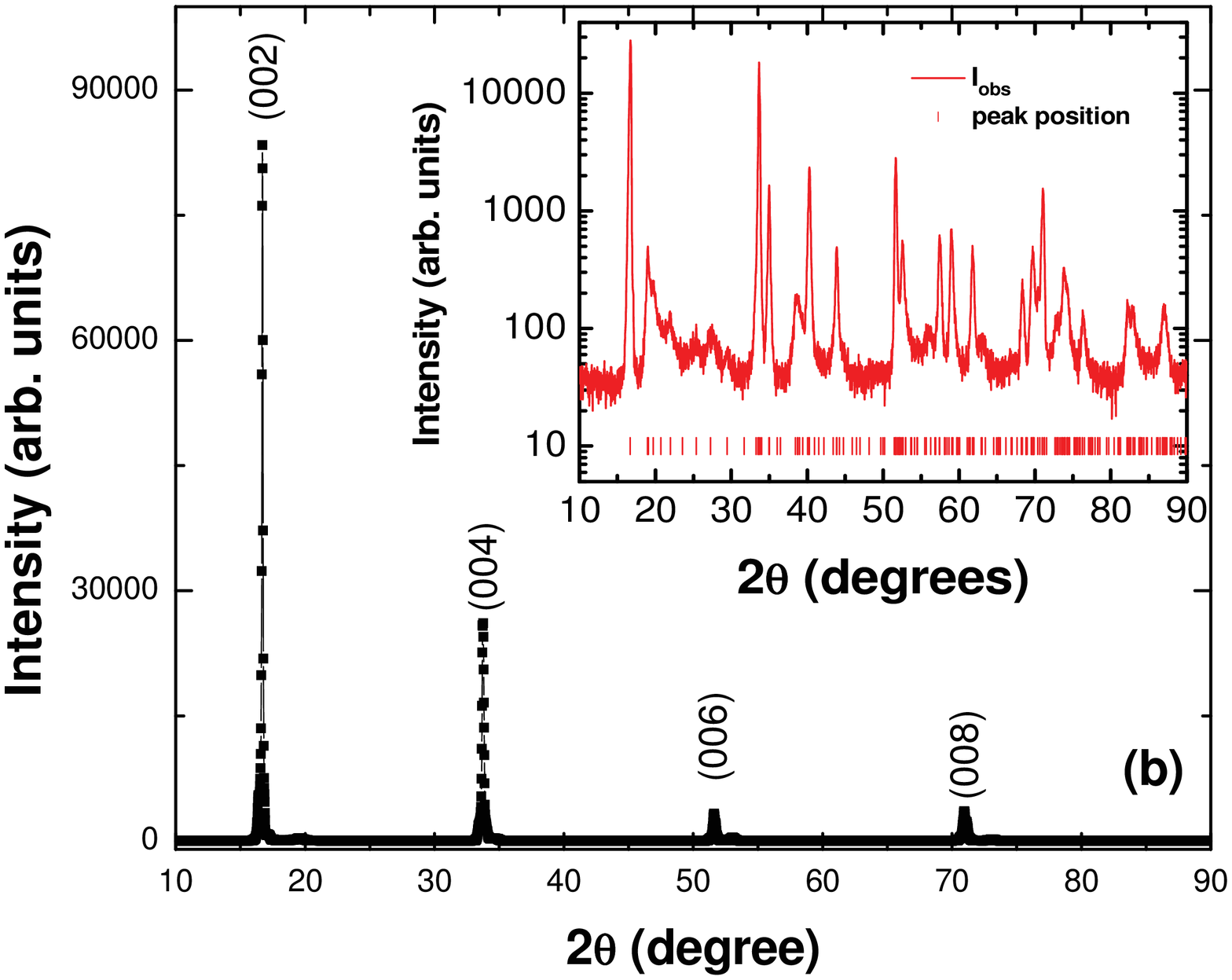}
\caption{(a) Rietveld refinement of the X-ray diffraction data of polycrystalline Na$_2$IrO$_3$ synthesized at 900~$^{\circ}$C\@.  The closed symbols represent the observed data, the solid lines represent the fitted pattern, the dotted lines represent the difference between the observed and calculated intensities and the vertical bars represent the peak positions.  (b) X-ray diffraction of $c$ axis oriented single crystals of Na$_2$IrO$_3$.  The inset shows the powder x-ray diffraction of crushed single crystals on a semi-log plot.   
\label{Fig-xrd}}
\end{figure}

\section{RESULTS}
\label{sec:RES}
\subsection{Crystal Structure and Chemical Analysis}
Powder x-ray diffraction (PXRD) scans of polycrystalline Na$_2$IrO$_3$ synthesized at 900~$^{\circ}$C are shown in Fig.\ref{Fig-xrd}(a).  All the lines in the PXRD pattern could be indexed to the monoclinic \emph{C2/c} (No.~15) structure.  Several $A_2T$O$_3$ ($A = $Li, Na, and $T = $Mn, Ru, Ir, Pd) type materials are known to adopt a similar structure.\cite{Breger2005,Kobayashi1995,Kobayashi2003,Panina2007}  The structure is made up of layers containing only the $A$ atoms alternating with $AT_2$O$_6$ layers stacked along the $c$ axis as shown in Fig.~\ref{Fig-structure}(a) for Na$_2$IrO$_3$.  Within the NaIr$_2$O$_6$ layers the edge sharing IrO$_6$ octahedra form a honeycomb lattice as shown in Fig.~\ref{Fig-structure}(b).  The Na atoms occupy voids between the IrO$_6$ octahedra.  It is known that these materials commonly posses a large amount of disorder arising from faults in the stacking of the $AT_2$O$_6$ layers.  This leads to anisotropic lineshapes in the X-ray diffraction patterns and reduced intensities of peaks between about $2\theta = 20$~--~$35 ^{\circ}$.\cite{Breger2005}  To account for the effect of this stacking disorder in Rietveld refinements of the powder X-ray data, site mixing between the $A$ and $T$ sites within the $AT_2$O$_6$ layers have been introduced.\cite{Breger2005,Kobayashi2003,Panina2007}  We have used a similar approach.  Rietveld refinements,\cite{Rietveld} shown in Fig.~\ref{Fig-xrd}, of the X-ray pattern gave the unit cell parameters $a$~=~~5.4198(5) \AA , $b$~=~9.3693(3) \AA\, $c$~=~10.7724(7) \AA\, and $\beta = 99.568(23)~^{\circ}$.  The fractional atomic positions, occupancies, isotropic thermal factors, and the reliability parameters $R_{\rm wp}$ and $R_{\rm p}$ obtained from the Rietveld refinement are given in Table~\ref{tabStruct}.  We find that a substantial ($\sim 15\%$) site mixing between Ir and Na within the NaIr$_2$O$_6$ layers is required to fit the x-ray patterns indicating a large amount of atomic or stacking disorder.  The introduction of site disorder allows the intensities of peaks between $2\theta = 20$~--~$35 ^{\circ}$ to be fit quite well, however, the peak profiles are still poorly fit.  This leads to large reliability factors as listed in Table~\ref{tabStruct}.  For a complete structural analysis, stacking faults have to be incorporated in the fits of the x-ray patterns as was done for example, for Li$_2$MnO$_3$.\cite{Breger2005}

\begin{table*}

\caption{\label{tabStruct}
Structure parameters for Na$_2$IrO$_3$ obtained for Rietveld refinements of powder XRD data.  The overall isotropic thermal parameter $B$ is defined within the temperature
factor of the intensity as $e^{-2B \sin^2 \theta/ \lambda^2}$.}
\begin{ruledtabular}
\begin{tabular}{|c||ccccccc|}
atom & \emph{x} & \emph{y} & \emph{z} & Occupancy&$B$ &$R_{\rm wp}$&$R_{\rm p}$\\
& & && &(\AA$^2$) & \\\hline  
Ir~~ & 0.2726(4) & 0.0781(3) & 0.0005(2) & 0.856(3)&0.008(2)& 0.212&0.165\\
Na~~ &  &  &  & 0.144(2)&0.013(5)&  & \\ \hline
Na~~ &0.75 & 0.25 & 0.0 & 0.735(2)&0.009(3) & & \\
Ir~~ & &  &  & 0.265(4)&0.007(1)& & \\ \hline
Na~~ &0.50 & 0.5792(1) & 0.25 & 1.0&0.012(7)&  &\\  
Na~~ &0.0 & 0.4720(4) & 0.25 & 1.0&0.009(2)& & \\
Na~~ &0.0 & 0.75 & 0.25 & 1.0&0.016(4)&  &\\  
O~~ &0.3795(3) & 0.2777(4) & -0.1002(1) & 1.0&0.013(8)&  &\\ 
O~~ &0.5966(5) & 0.0513(3) & 0.12128(4) & 1.0&0.013(2)&  &\\ 
O~~ &0.6579(4) & 0.4130(6) & 0.1086(2) & 1.0&0.014(6)&  &\\  
\end{tabular}
\end{ruledtabular}
\end{table*}
 
PXRD scans of crushed single crystals showed large preferred orientation along $c$ axis as expected for a layered material and peaks other than (00l) were much lower in intensity compared to the (00l) lines.  However, when plotted on a semi-log plot all the lines in the X-ray patterns could be indexed to the monoclinic \emph{C2/c} (No.~15) structure.  The PXRD data of crushed single crystals is shown in Fig.~\ref{Fig-xrd}(b) inset along with the peak positions expected for the monoclinic \emph{C2/c} (No.~15) structure.  Due to the large preferred orientation a Rietveld refinement of the PXRD pattern for crushed single crystals was not possible.  However, from fitting the peak positions we obtain the lattice parameters $a$~=~~5.4262(7) \AA , $b$~=~9.3858(6) \AA\, $c$~=~10.7688(6) \AA\, and $\beta = 99.580(15)~^{\circ}$.  These values are close to those obtained above for the polycrystalline material.   
The x-ray pattern of surfaces of the as-grown single crystals are shown in Fig.\ref{Fig-xrd}(b).  Only (00l) reflections expected for the monoclinic \emph{C2/c} (No.~15) structure are observed indicating that the surface of the crystals are perpendicular to the crystallographic $c$ axis.

An elemental analysis of the single crystals using a scanning electron microscope (SEM) showed the presence of only the desired elements Na, Ir, and O in the crystals and an energy dispersive x-ray analysis gave the Na : Ir ratio 1.88 : 1 which is close to the expected 2 : 1 ratio for Na$_2$IrO$_3$.   
 
\subsection{Electrical Resistivity}
\begin{figure}[t]
\includegraphics[width=2.7in, angle=-90]{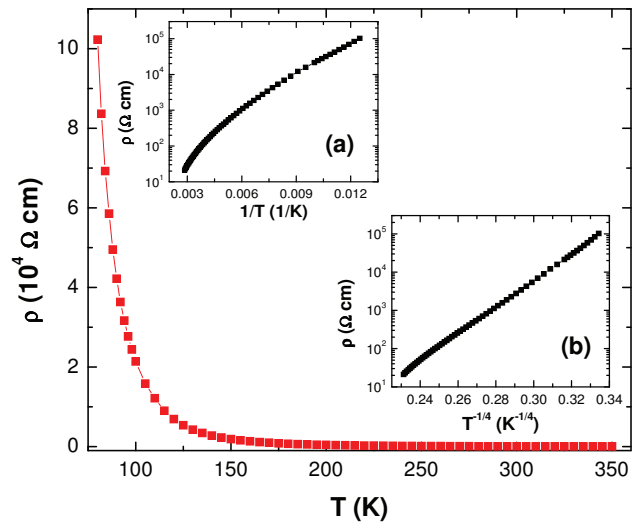}
\caption{(Color online) The in-plane electrical resistivity $\rho$ versus temperatures $T$ for a single crystal of Na$_2$IrO$_3$. The inset~(a) shows the $\rho$ versus $1/T$ data on a semi-log scale.  The inset~(b) shows the $\rho$ versus $1/T^{1/4}$ data on a semi-log scale.  
\label{Fig-res}}
\end{figure}
 
Figure~\ref{Fig-res} shows the in-plane dc electrical resistivity of a single crystal of Na$_2$IrO$_3$ between $T = 80$~K and 350~K\@.  The $T = 350$~K value $\rho(350 {\rm K}) \approx 21 \Omega$~cm and the temperature dependence strongly indicate that Na$_2$IrO$_3$ is an insulator.  The inset~(a) shows the $\rho$ versus $1/T$ data on a semi-log plot.  The data clearly does not follow an Arrhenius law and could not be fit with an activated behavior $\rho(T) \propto exp(-\Delta/T)$ in any extended range of temperature.  As shown in Fig.~\ref{Fig-res} inset~(b) however, the data follows a $\rho(T) \propto {\rm exp}[(\Delta/T)^{1/4}]$ behavior between 100~K and 300~K with deviations at higher and lower $T$.  Such a behavior is associated with three-dimensional variable-range-hopping of carriers localized by disorder.  The observation of such a $T$ dependence in Na$_2$IrO$_3$ is not understood at present.  A similar $T$ dependence was observed for single crystals of Sr$_2$IrO$_4$ in a limited $T$ range.\cite{Cao1998}

\subsection{Isothermal Magnetization}
\begin{figure}[t]
\includegraphics[width=2.5in, angle=-90]{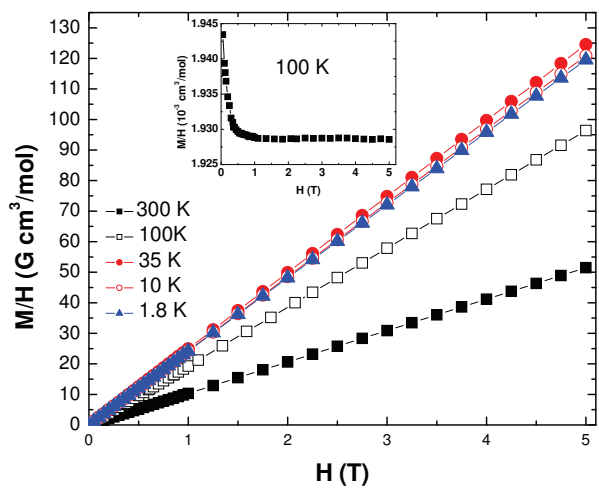}
\caption{(Color online) Isothermal magnetization $M$ versus magnetic field $H$ at various temperatures $T$ for a collection of randomly oriented single crystals of Na$_2$IrO$_3$.  
\label{Fig-M(H)}}
\end{figure}

To check for the presence of any ferromagnetic impurities we present in Fig.~\ref{Fig-M(H)} a selection of isothermal magnetization $M$ versus magnetic field $H$ data measured at various temperatures $T$ on a collection of randomly oriented crystals with total mass $m = 86.34$~mg.  The $M(H)$ data above $T = 50$~K are proportional to $H$ indicating the absence of any ferromagnetic impurities in the material.  The Fig.~\ref{Fig-M(H)} inset shows $M/H$ versus $H$ data at $T = 100$~K to demonstrate that above $\approx 5000$~G, $M/H$ is a constant independent of $H$.  At lower $T$ the $M(H)$ data begin to show curvature indicating the growing influence of magnetic interactions in the material.  The $T = 10$~K and $T = 1.8$~K $M(H)$ curve lie below the $M(H)$ curve at $T = 35$~K\@.  The reason for these behaviors is the presence of a long-range antiferromagnetic state at low $T$ in this material as discussed below.  

\subsection{Magnetic Susceptibility}
\begin{figure}[t]
\includegraphics[width=2.5in, angle=-90]{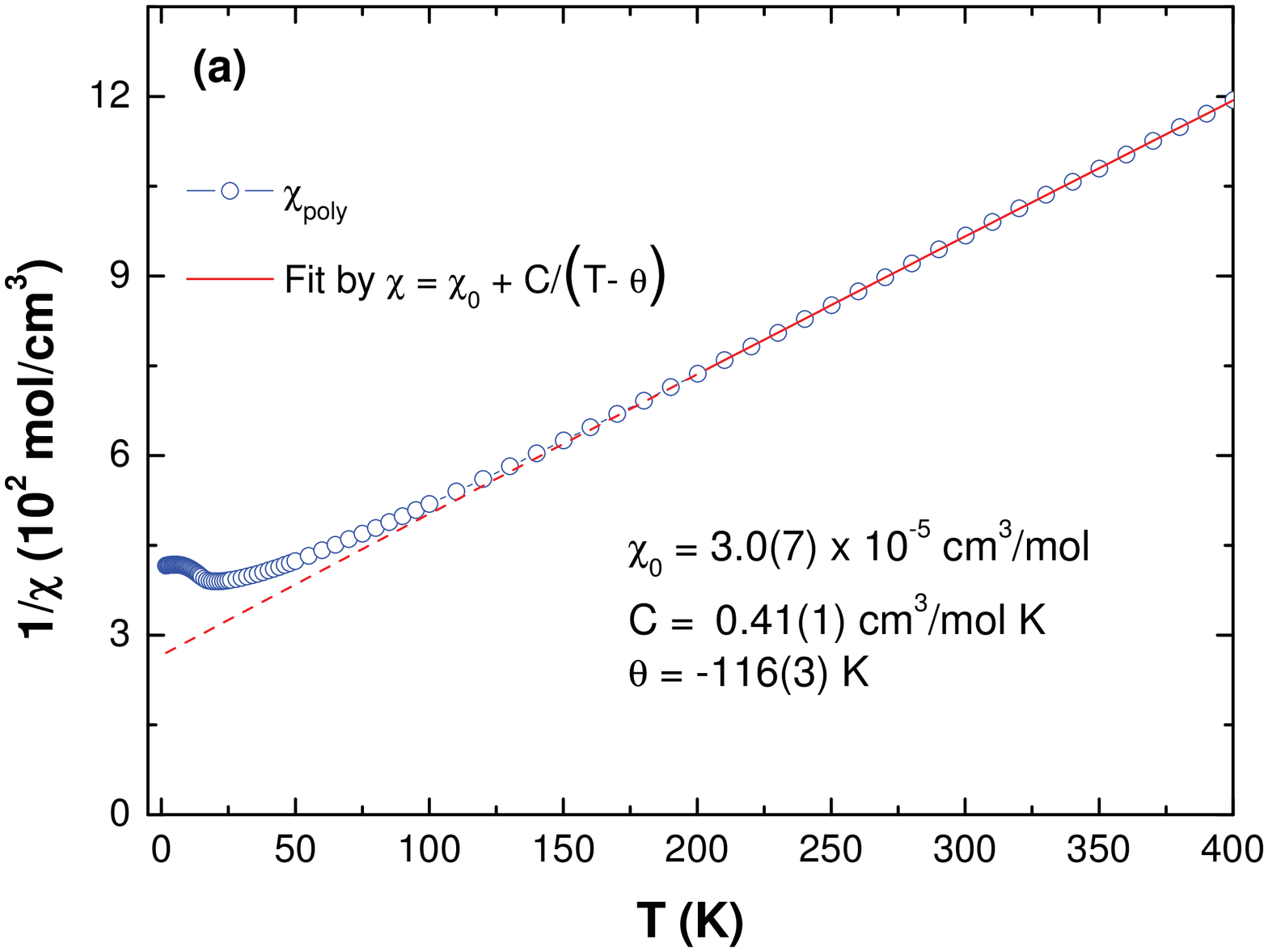}
\includegraphics[width=2.5in, angle=-90]{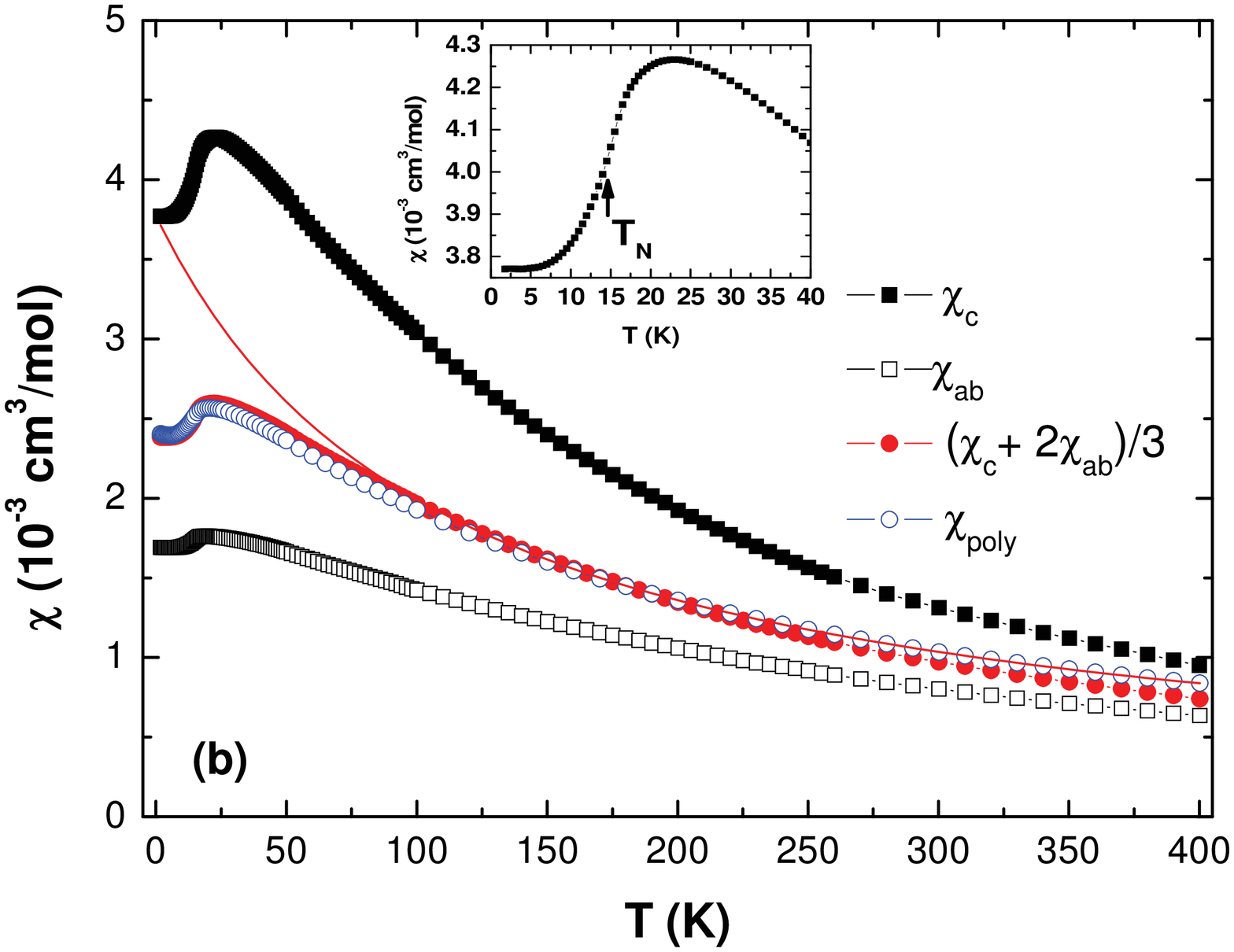}
\caption{(Color online) (a) Inverse magnetic susceptibility $1/\chi = H/M$ versus temperature $T$ for a collection of randomly oriented single crystals of Na$_2$IrO$_3$ in a magnetic field $H = 2$~T\@.  The solid curve through the data is a fit by the expression $\chi = \chi_0 + C/(T-\theta)$ and the dashed curve is an extrapolation to $T = 0$.  (b) the anisotropic magnetic susceptibilities $\chi_c$ and $\chi_{ab}$ versus $T$.  The powder average susceptibility $(\chi_c + 2\chi_{ab})/ 3$, and the polycrystalline susceptibility $\chi_{\rm poly}$ are also shown.  The solid curve through the $\chi_{\rm poly}$ data is the Curie-Wiess fit.  The inset shows the low $T$ $\chi_c(T)$ data to highlight the broad maximum at $T \approx 23$~K\@.  The arrow indicates the onset temperature $T_{\rm N} = 15$~K for the long-ranged antiferromagnetic ordering.   
\label{Fig-chi}}
\end{figure}

The inverse magnetic susceptibility $1/\chi_{\rm poly} = H/M$ versus $T$ data measured between $T = 1.8$~K and 400~K in an applied magnetic field $H = 2$~T for a collection of randomly oriented single crystals of Na$_2$IrO$_3$ are shown Fig.~\ref{Fig-chi}(a).  The $1/\chi_{\rm poly}(T)$ data between $T = 200$~K and 400~K were fit by the Curie-Wiess expression $\chi = \chi_0 + {C\over T-\theta}$ with $\chi_0$, $C$, and $\theta$ as fitting parameters.   The fit, shown in Fig.~\ref{Fig-chi}(a) as the solid curve through the data and extrapolated to $T = 0$, gave the values $\chi_0 = 3.0(7)\times10^{-5}$~cm$^3$/mol, $C = 0.41(1)$~cm$^3$~K/mol, and $\theta = - 116(3)$~K, respectively.  The above value of $C$ corresponds to an effective moment of $\mu_{\rm eff} = 1.81(2)~\mu_{\rm B}$ assuming a $g$-factor $g = 2$.  This value of $\mu_{\rm eff}$ is close to the value 1.74~$\mu_{\rm B}$ expected for spin~=~1/2 moments.  This indicates that the Ir$^{4+}$ moments are in an effective spin $S_{\rm eff} = 1/2$ state.  The large and negative $\theta = -116(3)$~K further indicates that strong antiferromagnetic interactions exist between these $S_{\rm eff} = 1/2$ moments.  The deviation of the $1/\chi(T)$ data from the Curie-Weiss expression below about $T = 100$~K is ascribed to the increasing influence of these antiferromagnetic correlations.  

The anisotropic magnetic susceptibilities are shown in Fig.~\ref{Fig-chi}(b).  $\chi_{\rm c}$ and $\chi_{\rm ab}$ are the magnetic susceptibilities measured with $H = 2$~T applied along the $c$ axis and perpendicular to the $c$ axis, respectively.  The susceptibility is weakly anisotropic with the out of plane $\chi_{\rm c}$ being larger than the in-plane $\chi_{\rm ab}$ in the whole $T$ range.   We find ${\chi_{\rm c}\over \chi_{\rm ab}} = 1.5 ~{\rm and}~ 2.3$ at $T = 400$~K and 1.8~K, respectively.  The anisotropy can originate from anisotropic van vleck paramagnetic susceptibility and/or from anisotropy in the $g$-factor which might both be expected if a trigonal crystal field is present.  If we assume $S = 1/2$ and fit the $\chi_c(T)$ and $\chi_{ab}(T)$ data to a Curie-Wiess behavior with $\theta = -116$~K obtained for the $\chi_{\rm poly}$ data, we obtain $g$-factors of $g_c = 2.68(3)$ and $g_{ab} = 1.87(2)$, respectively.  Although we have not included an anisotropic van Vleck term, the above analysis suggests that the $g$ factors might be highly anisotropic.  This can be confirmed by electron spin resonance measurements. 

The powder average susceptibility $\chi_{\rm powder} = {\chi_{\rm c} + 2 \chi_{\rm ab}\over 3}$ and $\chi_{\rm poly}$ [from Fig.~\ref{Fig-chi}(a)] are also shown in Fig.~\ref{Fig-chi}(b).  $\chi_{\rm powder}$ matches quite well with $\chi_{\rm poly}$ except at high $T$ where the small diamagnetic signal from the quartz holder lowers $\chi_{\rm powder}$.  

The Curie-Wiess (CW) fit obtained in Fig.~\ref{Fig-chi}(a) above is also shown as the solid line through the $\chi_{\rm poly}$ data in Fig.~\ref{Fig-chi}(b).  As mentioned before, the data deviate from the CW behavior below about $T = 100$~K and passes over a broad maximum at about 23~K before dropping abruptly below $T \approx 15$~K\@.  This can be seen from Fig.~\ref{Fig-chi}(b) inset which shows the $\chi_{\rm c}(T)$ data below $T = 40$~K on an expanded scale.  The sharp drop below $T_{\rm N} \approx 15$~K is associated with the onset of long-ranged antiferromagnetic ordering in Na$_2$IrO$_3$ while the broad maximum above the ordering is most likely associated with short-ranged order seen commonly in low-dimensional magnetic materials.  This is supported by our heat capacity measurements presented below.

\subsection{Heat Capacity}
Figure~\ref{Fig-heatcap}(a) shows the heat capacity divided by temperature $C/T$ versus temperature $T$ data measured between $T = 2$~K and 40~K in a zero applied magnetic field $H$.  The heat capacity of a polycrystalline sample of Na$_2$SnO$_3$ is also shown in the same figure as an approximate estimation of the lattice contribution to the heat capacity of Na$_2$IrO$_3$.  The lambda-like anomaly at $T_{\rm N} = 15$~K for Na$_2$IrO$_3$ confirms the bulk nature of the antiferromagnetic ordering observed in the $\chi$ data in Fig.~\ref{Fig-chi} above.  Figure~\ref{Fig-heatcap}(a) inset shows the $C/T$ versus $T$ data between $T = 12$~K and 19~K, measured in $H = 0$ and $H = 7$~T\@.  The slight depression of $T_{\rm N}$ in an applied magnetic field indicates the antiferromagnetic nature of the ordering.     

\begin{figure}[t]
\includegraphics[width=2.5in, angle=-90]{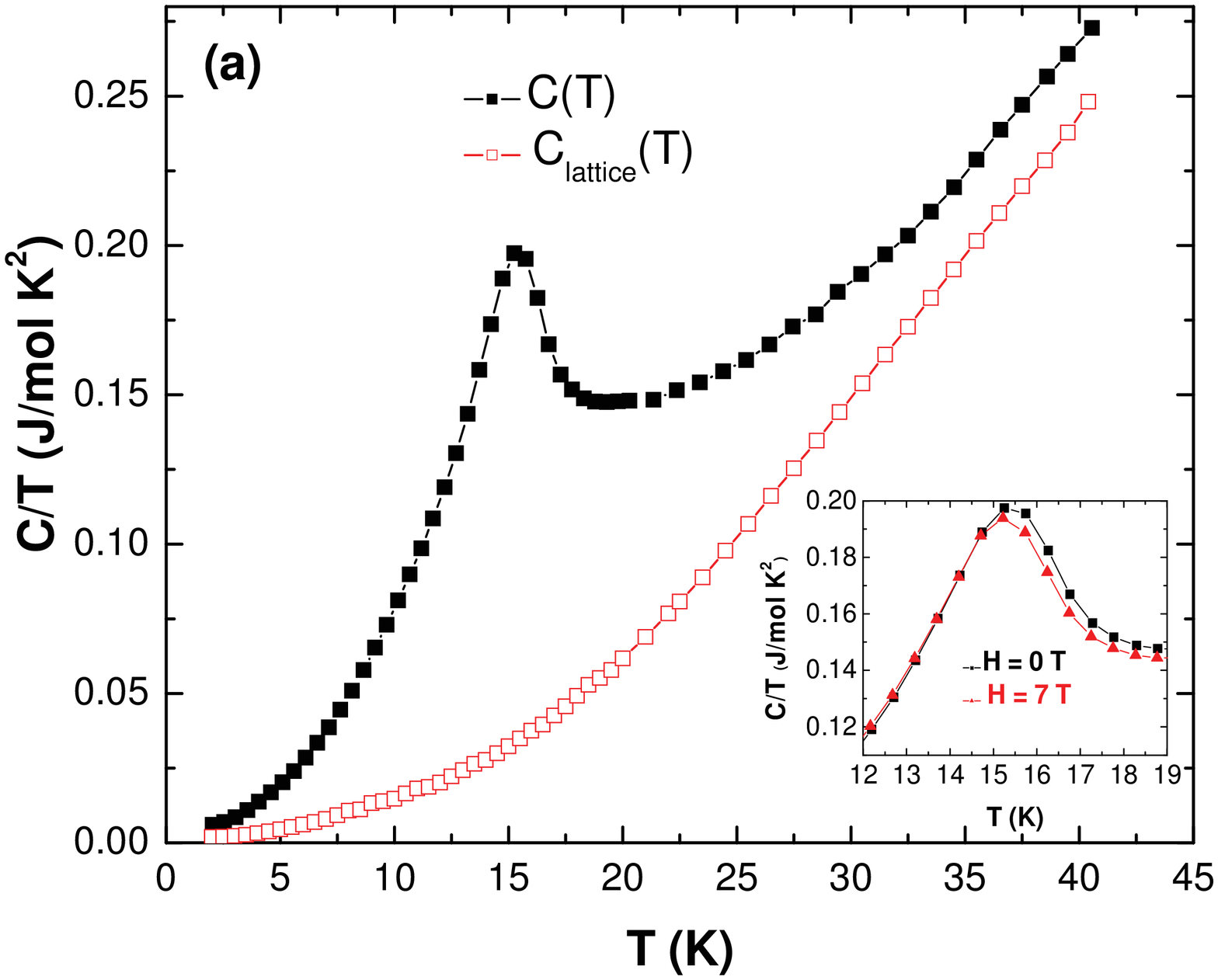}
\includegraphics[width=2.5in, angle=-90]{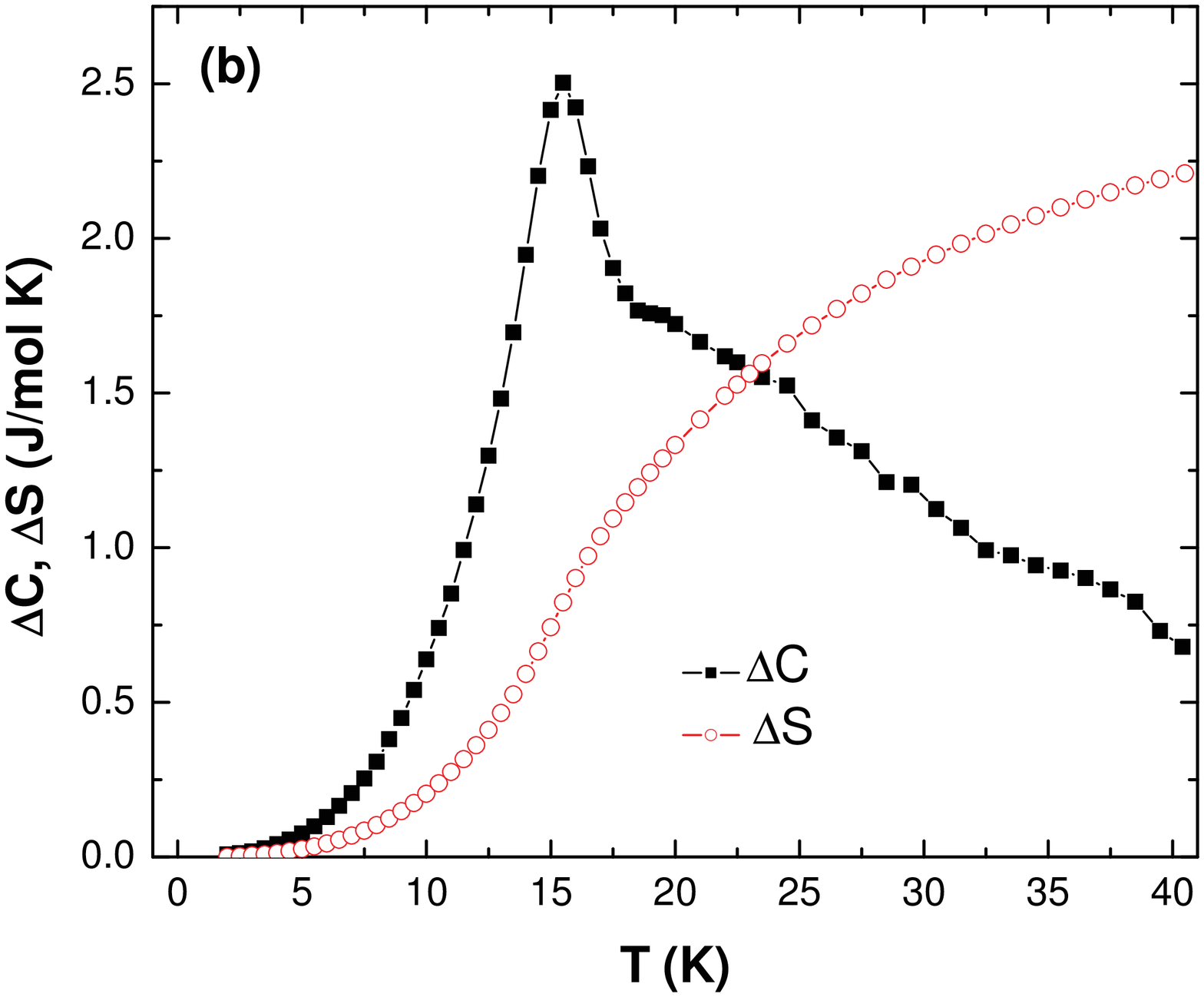}
\caption{(Color online) (a) The heat capacity divided by temperature $C/T$ versus $T$ data between $T = 1.8$~K and 40~K for single crystals of Na$_2$IrO$_3$ and the heat capacity of Na$_2$SnO$_3$ as the lattice contribution $C_{\rm lattice}/T$ versus $T$.  The inset shows the $C/T$ versus $T$ data in $H = 0$ and 7~T applied magnetic field.  (b)  The difference heat capacity $\Delta C$ and difference entropy $\Delta S$ versus $T$ data between $T = 1.8$~K and 40~K\@.    
\label{Fig-heatcap}}
\end{figure}

Figure~\ref{Fig-heatcap}(b) shows the difference heat capacity $\Delta C(T) = C(T) - C_{\rm lattice}(T)$ and the difference entropy $\Delta S(T)$ obtained by integrating the $\Delta C(T)/T$ versus $T$ data.  The $\Delta C$ data shows a sharp peak at $T_{\rm N} = 15$~K and a broad tail which extends to higher $T$.  This suggests the presence of short-ranged order above the bulk three-dimensional ordering which occurs at $T_{\rm N}$.  This is supported by the fact that the entropy just above $T_{\rm N}$ is $\Delta S(17 {\rm K}) \approx 1.2$~J/mol~K which is only about 20\% of the value Rln(2)~=~5.76~J/mol~K expected for ordering of $S = 1/2$ moments.  $\Delta S$ also continues to increase upto the highest $T$ of our measurements.  

\section{Summary and Discussion}
Single crystals of Na$_2$IrO$_3$ have been grown and their structural, electrical transport, magnetic, and thermal properties investigated.  Na$_2$IrO$_3$ possesses a layered structure where pure Na layers are stacked alternately with NaIr$_2$O$_6$ slabs along the $c$ axis of the monoclinic unit cell.  Within the NaIr$_2$O$_6$ layers, the Ir$^{4+}$ moments sit on a hexagonal lattice.  Electrical transport within the $ab$ plane shows insulating behavior.  Magnetic susceptibility measurements provide evidence that the Ir atoms carry effective $S_{\rm eff} = 1/2$ moments and they have strong antiferromagnetic interactions as evidenced by a large and negative Weiss temperature $\theta = -116(3)$~K\@.  The magnetic susceptibility is slightly anisotropic with $\chi_{\rm c}$ being the easy axis of magnetization.  

The $\chi(T)$ deviates from Curie-Wiess behavior below $T \sim 100$~K and passes over a broad maximum around $T = 23$~K before decreasing strongly below the three-dimensional antiferromagnetic ordering temperature $T_{\rm N} = 15$~K\@.  At the lowest $T = 1.8$~K, $\chi(T)$ saturates to a large and finite value.  This indicates that the magnetic ordering is most likely non-collinear.  It is difficult to decide the ordering direction from our measurements.  The reduction in the absolute value of $\chi$ is larger for $\chi_c$ but the change of slope at $T_{\rm N}$ is stronger for $\chi_{ab}$. 

A much smaller $T_{\rm N}$ compared to the Wiess temperature $\theta= -116(3)$~K indicates frustrated magnetic interactions in Na$_2$IrO$_3$.  This is consistent with a recent theoretical study which has made estimates of the ratio between the nearest-neighbor (NN) and next-nearest-neighbor (NNN) magnetic interactions and find that ${J_{\rm NNN}\over J_{\rm NN}} = 0.47$, indicating strongly frustrated magnetic interactions.\cite{Jin2009}  The maximum in $\chi(T)$ above the long-ranged ordering temperature $T_{\rm N}$ is commonly observed in low dimensional materials where short ranged magnetic order develops (within the NaIr$_2$O$_6$ layers in Na$_2$IrO$_3$ for example) well above $T_{\rm N}$ and long-ranged magnetic order only occurs when the interplanar interactions become important.  The presence of short-ranged magnetic order is also supported by our heat capacity measurements.  The $C(T)$ shows a sharp lambda-like anomaly at $T_{\rm N} = 15$~K indicating bulk magnetic ordering.  However, the difference heat capacity $\Delta C(T)$ shows a broad tail extending to much higher $T$ above $T_{\rm N}$ and the difference entropy $\Delta S$ just above $T_{\rm N}$ is only about 20\% of what is expected for ordering of spin $S = 1/2$ moments. The $\Delta S(T)$ also continues to increase upto the highest $T$ of our measurements.  An incorrect estimation of the lattice contribution to $C(T)$ can lead to a reduced $\Delta S(T)$.  However, if we estimate the entropy by integrating the total $C(T)/T$ versus $T$ data without any lattice subtraction, we still end up with only $\approx$~30\%Rln(2) at $T_{\rm N}$ suggesting that the reduced entropy is intrinsic to Na$_2$IrO$_3$ and most likely occurs due to the developement of short-ranged magnetic order well above $T_{\rm N}$.      

There has been a lot of recent theoretical interest in Na$_2$IrO$_3$ with suggestions of a topological band insulator\cite{Shitade2009} or a magnetically ordered Mott insulator\cite{Jin2009}  being possible ground states for this system.  Our electrical transport results indicate insulating behavior consistent with the above predictions.  However, our magnetic measurements indicate local moment magnetism and long-range antiferromagnetic ordering which is not consistent with the topological band insulator picture above.

Further experiments will be needed to investigate the importance of spin-orbit interactions and the strength of electronic correlations in Na$_2$IrO$_3$ as has been suggested in previous theoretical work.\cite{Shitade2009,Jin2009}  The magnetic structure and the nature of the magnetic interactions will also need further work to compare with the various proposed magnetic ground states although the spin-liquid phase near the Kitaev limit can be ruled out from our results .\cite{Chaloupka2010}  

To summarize, single crystals of Na$_2$IrO$_3$ have been synthesized.  From magnetic, thermal, and transport measurements we conclude that this is a quasi-low-dimensional, magnetically frustrated material which undergoes long-ranged antiferromagnetic ordering below $T_{\rm N} = 15$~K into most likely a non-collinear magnetic structure.  The in-plane electrical transport properties together with the local moment magnetism indicate that Na$_2$IrO$_3$ is a Mott insulator.  

\begin{acknowledgments}
Y.S. acknowledges corrspondence and discussion with G. Khaliullin.  Y.S. would like to thank the Alexander von Humboldt foundation for a research fellowship.  \\
\end{acknowledgments}

\end{document}